\documentclass[12pt]{article}
\usepackage{times}
\usepackage{geometry}
\usepackage[utf8]{inputenc}
\usepackage{enumitem,amssymb}
\usepackage{ragged2e}
\newlist{thematic}{itemize}{8}
\setlist[thematic]{label=$\square$}
\usepackage{pifont}
\usepackage{hyperref}
\usepackage{graphicx}
\usepackage{amsmath}
\newcommand{\cmark}{\ding{51}}%
\newcommand{\done}{\rlap{$\square$}{\raisebox{2pt}
\def\ltsim{\mathrel{\hbox{\rlap{\hbox{\lower3pt\hbox{$\sim$}}}\hbox{$<$}}}}
\def\gtsim{\mathrel{\hbox{\rlap{\hbox{\lower3pt\hbox{$\sim$}}}\hbox{$>$}}}}
{\large\hspace{1pt}\cmark}}%
\hspace{-2.5pt}}

\usepackage{xcolor}

\def\gaia{{\em Gaia}}

\begin{document}
\raggedright
\huge
Roman CCS White Paper \linebreak

New Compact Object Binary Populations with Precision Astrometry \linebreak
\normalsize

\noindent\textbf{Roman Core Community Survey:} Galactic Bulge Time Domain Survey (GBTDS)

\noindent\textbf{Scientific Categories:} Stellar physics and stellar types

\noindent\textbf{Additional scientific keywords:} Astrometry, Compact Objects, Black holes, Neutron stars, Binary stars
\newline

\textbf{Submitting Author:}

Name:	P. Gandhi
 \linebreak						
Affiliation:   University of Southampton, Highfield SO17\,1BJ, UK
 \linebreak
Email:    poshak.gandhi@soton.ac.uk
 \linebreak
 
\textbf{Contributing authors:}
  \linebreak
  C.\,Dashwood Brown (Univ. Southampton)\\
  Y.\,Zhao (Univ. Southampton)\\
  K.\,El-Badry (Harvard Univ. CfA)\\
  T.J.\,Maccarone (Texas Tech)\\
  C.\,Knigge (Univ. Southampton)\\
  J.\,Anderson (STScI)\\
  M.\,Middleton (Univ. Southampton)\\
  J.C.A.\,Miller-Jones (ICRAR, Curtin University)
  \vspace{1cm}

\textbf{Abstract:} 
Compact object binaries (a black hole or a neutron star orbiting a non-degenerate stellar companion) are key to our understanding of late massive star evolution, in addition to being some of the best probes of extreme gravity and accretion physics. \gaia\ has opened the door to astrometric studies of these systems, enabling geometric distance measurements, kinematic estimation, and the ability to find new previously unknown systems through measurement of binary orbital elements. Particularly puzzling are newly found massive black holes in wide orbits ($\sim$\,AU or more) whose evolutionary history is difficult to explain. Astrometric identification of such binaries is challenging for \gaia, with only two such examples currently known. Roman's enormous grasp, superb sensitivity, sharp PSF and controlled survey strategy can prove to be a game-changer in this field, extending astrometric studies of compact object binaries several mag deeper than \gaia. We propose to use the microlensing Galactic Bulge Time Domain Survey to identify new wide-orbit black hole compact object binaries, determine their prevalence and their spatial distribution, thus opening up new parameter space in binary population studies.

\pagebreak

\section{Introduction}

Most massive stars exist in binaries, currently the best systems for identifying compact objects from the X-ray power they emit during accretion \cite{casares14}. The more massive compact progenitor evolves and interacts with the stellar companion, either via Roche lobe overflow (where the star has expanded beyond its Roche lobe, resulting in mass transfer onto the compact object) or through a common envelope phase (where both stellar cores orbiting within the expanding envelope of the mass-losing supergiant). The compact object progenitor continues to evolve and eventually undergoes a supernova, forming a neutron star or black hole. The evolution of these binary systems is complex and many factors remain poorly constrained, including the impact of natal kicks, of metallicity on massive star wind loss, and the impact of common envelope evolution, amongst others \cite{verbuntvandenheuvel95, PortegiesZwart97, nelemans99, belczynski20}. The more massive amongst these binaries are potential progenitors for the LIGO/Virgo/Kagra gravitational wave [GW] population, and factors such as natal kicks can dramatically change predictions for GW merger rates \cite{belczynski20}. Understanding these factors is thus critical and can be done by detailed follow-up of known X-ray binaries with electromagnetic follow-up and precision astrometry \cite{Blaauw61, Brandt95, Hobbs05, giacobbo18, Mirabel17, Gandhi19, Atri19,ODoherty23}. 

\hspace{1cm}But our knowledge of compact objects in the Milky Way remains strongly incomplete. For instance, estimates of the number of accreting compact objects in binary systems with a donor star range over 10$^{3-8}$ \cite{Pfahl03, Corral-Santana16}. About 500 candidates are known, mostly discovered from their X-ray emission when accreting material from a companion star, with only $\sim$\,30 dynamically confirmed as hosting a black hole [BH] \cite{Corral-Santana16}. Studies suggest that there may be an even larger populations of {\em non-interacting} compact object binaries with no ongoing accretion activity \cite{andrews19, chawla22}.

\begin{figure}
    \centering
    \includegraphics[angle=0,width=0.8\textwidth]{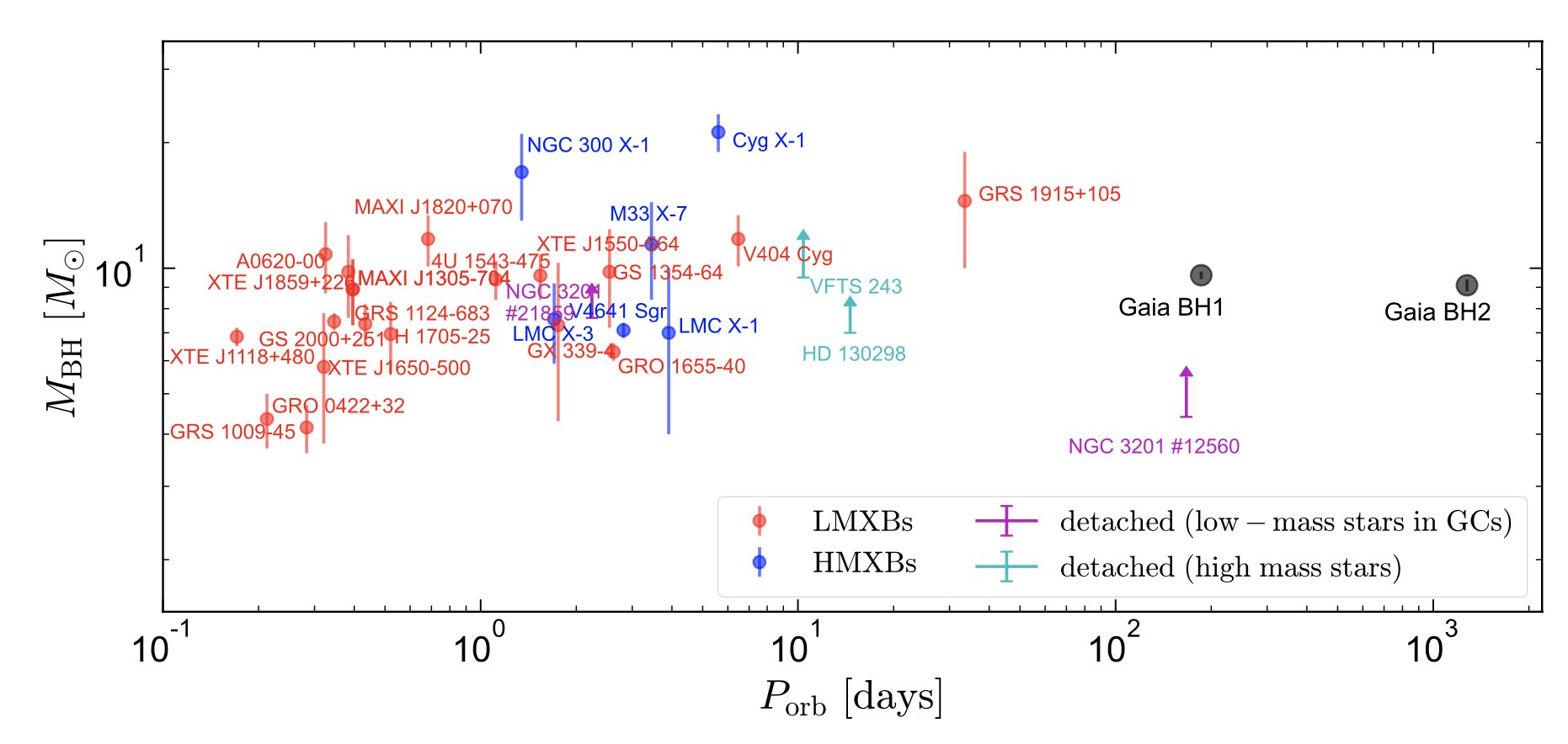}
    \caption{\textcolor{teal}{Mass vs. Orbital Period for dynamically confirmed black holes in X-ray binaries \cite{Corral-Santana16} and for the two recently discovered ultra-wide systesm Gaia-BH1 and BH2. These may be the tip of the iceberg of the wide-orbit black hole binary population. Figure from \cite{El-Badry23b}.}}
    \label{fig:wobble}
\end{figure}

\hspace{1cm}Recently, a handful of candidate non-interacting systems have been discovered, though the veracity of some remains under contention \cite{thompson19,bodensteiner22}. Amongst these are two black hole systems discovered astrometrically from the \gaia\ DR3 non-single star solutions \cite{gaiadr3_nss}, which allow measurement of the binary orbital parameters leading to mass estimation. These systems, Gaia-BH1 \cite{El-Badry23a} and Gaia-BH2 \cite{Tanikawa2023,El-Badry23b}, lie in extremely wide orbits (period $P$\,=\,186\,d and 1,277\,d, corresponding to projected semi-major axes of $a$\,$\approx$\,1.4\,AU and 5\,AU, respectively). These are at least an order-of-magnitude wider than most known X-ray binaries (Fig.\,\ref{fig:wobble}). Optical spectroscopy of the companion places direct constraints on the black hole mass in each case.

\hspace*{1cm}Simulations suggest that wide period systems must have received small natal kicks ($\sim$\,a few--a few 10s of km\,s$^{-1}$) and ought to preferentially exist with low mass-ratios (i.e. the ratio of the companion to compact object mass $M_2$/$M_1$\,$<$\,5). By contrast, the nominal mass ratio of Gaia-BH1 is $\sim$10, so this system is difficult to explain based on current understanding of binary evolution. Furthermore, the runaway motion of Gaia-BH1 ($\gtrsim 70\,\mathrm{km~s^{-1}}$ in excess of Galactic rotation) is also hard to reconcile with a low natal kick (Zhao et al., in prep.).

\hspace*{1cm}‘Standard’ formation channels would predict that progenitors of Gaia-BH1 would have either shrunk to smaller orbits during a common envelope phase, or have merged. Alternatively, if the progenitor binary had a very large initial period $>$3000\ days, then it may have evolved without interaction -- a channel known as ‘direct-supernova’ evolution \cite{Kalogera1998}. But eventually the primary undergoes a core-collapse supernova, and population synthesis studies suggest the majority ($>$80\%) of these binaries will be disrupted by the supernova. Gaia-BH2, with its longer orbit, is even more difficult to explain under canonical formation channels. These systems highlight how much remains to be understood about binary evolution -- whether the system interacts prior to supernova or not cannot be determined from the available data. The observed masses of the binary components seems more akin to the simulated systems that do not experience a common envelope phase, and one might naively assume this is a smoking gun for that evolutionary pathway. However, neither the kinematics nor the system parameters (e.g. eccentricity) can be easily reproduced in either scenario.

\hspace*{1cm}These systems could be the tip of the iceberg of a new underlying population just starting to be probed with precision astrometry. Identifying more wide-period binaries will greatly improve our knowledge of binary evolution, including constraining the role and distribution of natal kicks imparted on compact objects, and the preferred evolutionary channels prior to supernova. This is what we propose to do with Roman.

\section{How the capabilities of a Roman Survey will uniquely enable the investigation}

Our assumed survey and astrometric precision baseline parameters are listed in Table\,\ref{tab:tab}, and are informed by the expectations for the Galactic Bulge microlensing survey \cite{roman_microlens}, together with considerations of systematic uncertainties that may impact the overall final accuracy \cite{wfirst_astrometry_wp, roman_technical_report}. 

\begin{table}[h]
    \centering
    \caption{\textcolor{teal}{Baseline astrometric precision and accuracy expectations for Roman observations.}}
    \begin{tabular}{l|r}
    \hline
    {\bf Objective} & {\bf Expected performance}\\
    \hline
       Detector Plate scale & 110\,mas \\
       PSF (F146) & 105\,mas \\
       Best single-exposure centroiding precision  & 0.01 pixels\,$\equiv$\,1.1\,mas \\
       Final parallax precision & $\approx$\,0.005\,mas\\ 
       Seasonal astrometric precision ($\sigma_{\rm ast}$) & $\approx$\,0.013\,mas\\ 
       Absolute a posteriori accuracy, from fitting \gaia\ field stars  & $\approx$\,0.05\,mas\\
       Photometric precision ($\sigma_{\rm F146,\, phot}$) & 0.01  @$m_{\rm AB}$\,$\approx$\,23 (AB)\\
       Sky coverage ($A$) & 2\,deg$^2$\\
       Number of field stars & 100\,million @$m_{\rm AB}$\,$\approx$\,23 (AB)\\
       Cadence & 15\,min for 72\,days ($\times$\,6 seasons)\\
    \hline
    \end{tabular}
    \label{tab:tab}
\end{table}

\hspace*{1cm}The critical criterion for finding compact objects in binaries is high photometric precision. At the distance of the Galactic centre $d$\,=\,8\,kpc, a star in a binary orbit with semi-major axis $a$\,=\,1\,AU (say) will display an angular wobble $\omega$\,=\,0.125 mas. In order to trace this orbit astrometrically, the demanded positional accuracy must be a small fraction of this. 

\hspace*{1cm}Orbital tracing also requires relatively high observational cadence. Gaia-BH1, for instance, has a period $P$\,$\approx$\,6\,months. Furthermore, knowing the source distance allows inference of the binary physical separation $a$ from the angular measurements. This, in turn, requires robust parallax estimation. 

\hspace*{1cm}The proposed framework of the Roman microlensing GBTDS can fulfil these criteria. The final predicted parallax precision (Table\,\ref{tab:tab}) is more than an order of magnitude better than that needed for estimating source distances out to the Galactic centre. The precision with which orbits can be traced can be approximated by taking the {\em seasonal} data as the relevant baseline. We assume a best single-exposure angular precision of 0.01\,pixel, or about 1.1\,mas \cite{wfirst_astrometry_wp}, and a subsequent gain of $\sqrt{N_{\rm exp}}$. Over a 72\,day season, the number of exposures gathered will be $N_{\rm exp}$\,=\,6912, resulting in a gain factor of 83, i.e. a seasonal angular precision of $\sigma_{\rm ast}$\,=\,0.013\,mas, yielding enough resolution for wobble estimation. 

\section{Figure of Merit}

A detailed selection function of compact object binaries will require knowledge of the space density of objects like Gaia-BH1, which remains unknown. For the purpose of planning Roman observations, we define a straightforward, albeit simplistic, figure-of-merit (FoM) here as {\bf Figure of Merit: Number of successful compact object binary identifications.} 

\begin{equation}
    {\rm FoM} = N \times {\rm A} \times \frac{1}{\sigma_{\rm ast}^p} \times \frac{1}{\sigma_{M_2}},
    \label{eq:fom}
\end{equation}

with $N$ being a normalisation factor which we ignore for the purposes of this discussion. The astrometric precision enters Eq.\,\ref{eq:fom} as $\sigma_{\rm ast}^p$. The value of the exponent $p$ will depend upon the spatial density of the targets of interest. Massive compact objects are very likely to be enhanced in the Galactic centre plane and bulge regions relative to the Solar neighbourhood, implying that their spatial density will increase with distance from Earth. Measurement of the orbital elements (including eccentricity and wobble) places further demands on astrometric precision, so $p$ is expected to be a positive value greater than 1. We here propose $p$\,=\,3 for FoM optimisation, based on the ansatz that the spatial density in the Galactic disc likely scales down as distance from the Galactic centre to the second power, and we require one additional exponent power for orbital element estimation. Higher $p$ values will simply give this factor greater weight. 

\hspace*{1cm}The final factor in the FoM denotes uncertainty estimates on the luminous companion star mass ($M_2$). This is required to translate the astrometric mass {\em function}, which places only a lower limit on the compact object mass $M_1$, to a direct constraint on $M_1$ \cite{andrews19}. Stellar mass constraints will be non-trivial, with the most robust constraints coming from future spectroscopic follow-up with facilities like the Extremely Large Telescope, or through deep space-based exposures. Initial estimates could be based on colour-magnitude analysis from multi-band photometry, so here we propose 

\begin{equation}
    \frac{1}{\sigma_{M_2}}\,=\,\sqrt{\sum_{f}{\frac{1}{\sigma_{f,\,{\rm phot}}^{2}}}}
\end{equation}

This equates the mass determination accuracy to the root mean square (r.m.s) photometric accuracy in all filters ($f$) observed, again ignoring any normalisation factors. GBTDS is likely to observe in at least two, and probably more, filters, allowing initial estimates of $\sigma_{M_2}$, and thus FoM.

\section{Observational strategy considerations} 

The fiducial proposed strategy of the microlensing survey would satisfy the requirements for uncovering the compact object binaries that we are interested in. But the strategy is not final, and we briefly outline how changing the survey parameters impact deliverables. 

\hspace{1cm}Whilst \gaia\ has been a superb astrometric machine, its observations are not custom tailored in any specific way. This means that there remain many systematic uncertainties in the data, resulting in a complex selection function \cite{castroginard23}. Upcoming \gaia\ data releases (DR4, and potentially, DR5) will likely reveal more of the sources that we seek, but the selection function is still likely to remain plagued with similar uncertainties. 

\hspace{1cm}By contrast, the proposed GBTDS focus on a limited set of fields, the huge grasp and direct imaging capabilities should yield a superior selection function, facilitating a range of population studies that are difficult with \gaia. The most critical requirement for optimising the FoM is, without doubt, knowledge of the PSF centroiding precision. But maintaining the PSF stability across observational seasons would boost FoM while also aiding in the selection function definition. Photometric precision plays a secondary, albeit important role. 

\hspace{1cm}Pushing to more sensitive mag limits would allow probing deeper into the bulge, which is likely to be rich in compact objects. But there would need to be a compromise between increased crowding and the extremes demands of astrometry at deeper limits. Understanding the impact of crowding will be important for our science. Expanding the survey to new areas (larger $A$ in Eq.\,\ref{eq:fom}) could instead prove more beneficial.

\hspace{1cm}More exposure time weighting to the shorter wavelength filters will provide a gain in terms of sharper PSF localisation. There would be a trade-off with the increased impact of Galactic extinction and our preferred filters would be F087 or longer. 

\hspace{1cm}Finally, a longer duration survey in future years will allow sampling of the upper end of the period distribution. Gaia-BH1's orbital period is about 6 months. Gaia-BH2, by contrast, has period $P$\,=\,1,277 days, so the GBTDS in its proposed incarnation (Table\,1) cover only half an orbit. With enough signal-to-noise, it is still possible to find such long period systems, though their characterisation will not be as robust.

\section{Ancillary Science}
The microlensing survey will uncover not only exoplanets, but also free-floating compact object lenses. Only a single isolated BH has so far been confirmed \cite{sahu22}. These isolated systems could have been born in binaries, but ejected following supernova. There are almost no firm constraints on connections between the isolated and bound compact object populations at this time. Expanding the samples of both will address this. 

\hspace{1cm}Furthermore, self-lensing binaries, in which an edge-on inclination results in photometric lensing of the companion star at superior conjunction will also be uncovered \cite{wiktorowicz21}. This is an independent technique for identifying compact objects, and will thus be free of any systematic biases impacting astrometric identification. Approximately $\sim$\,1 in 100 compact object binary systems will lie at an edge-on inclination angle as seen from Roman, allowing the possibility of self-lensing. Combining the self-lensing with the astrometric constraints for these should yield unprecedented, independent constraints on the system and component parameters. 

\hspace{1cm}Finally, new quiescent accreting systems could also be directly identified from their photometric and astrometric properties \cite{Gandhi22}. The population of accreting systems is also expected to rise towards the Galactic Bulge \cite{grimm02}, and proper characterisation of such systems with Roman will enable removing of contaminants such as active and variable stars and cataclysmic variables.

\bibliographystyle{abbrv}
\bibliography{romanwp}

\end{document}